\documentstyle[12pt]{article}

\input{psfig}

\font\mybb=msbm10 at 11pt
\def\bb#1{\hbox{\mybb#1}}

\newlength{\extraspace}
\newlength{\extraspaces}
\begin{document}

\addtolength{\baselineskip}{.8mm}

\thispagestyle{empty}

\begin{flushright}
{\sc OUTP}-96-68P\\
 \date \\
 hep-th/yymmxxx
\end{flushright}
\vspace{.3cm}

\begin{center}
{\large\sc{String Winding Modes From Charge Non-Conservation\\
in Compact Chern-Simons Theory}}\\[15mm]

{\sc  Leith Cooper,\footnote{leith@thphys.ox.ac.uk} 
Ian  I. Kogan\footnote{kogan@thphys.ox.ac.uk}
and Kai-Ming Lee\footnote{kmlee@sun1.phy.cuhk.edu.hk. 
Present address: Department of Physics, The Chinese University 
of Hong Kong, Shatin, N.T., Hong Kong.}} \\[2mm]
{\it Theoretical Physics, 1 Keble Road,
       Oxford, OX1 3NP, UK} \\[15mm]

{\sc Abstract}

\begin{center}
\begin{minipage}{14cm}
 
In this letter we show how string winding modes can
be constructed using topological membranes. We use 
the fact that monopole-instantons in compact topologically
massive gauge theory lead to charge non-conservation inside
the membrane which, in turn, enables us to construct vertex
operators with different left and right momenta. The amount of
charge non-conservation inside the membrane is interpreted as 
giving the momentum associated with the string winding mode and 
is shown to match precisely the full mass spectrum of compactified
string theory.

\end{minipage}
\end{center}

\end{center}

\noindent

\vfill
\newpage
\pagestyle{plain}
\setcounter{page}{1}
\renewcommand{\footnotesize}{\small}

Space-time compactification is of central importance to string theory:
it provides a mechanism for reducing higher-dimensional string theory 
down to four dimensions and is an alternative to the Chan-Paton method
for introducing isospin. Compactification, therefore, is a key to 
performing meaningful four-dimensional string phenomenology.
Space-time compactification also leads to the appearance of certain 
dualities \cite{duality} between string theories at strong and weak 
coupling and
therefore any new insight into the compactification mechanism 
may be helpful in finding a non-perturbative description of string 
theory. 

In this letter we examine space-time compactification 
in topological membrane ({\sc tm}) theory 
\cite{tmtheory}.
In particular, we consider the simple example of string 
theory compactified on a circle, which can be described by a compact 
$U(1)$ topologically massive gauge theory ({\sc tmgt}) \cite{tmgt}. 
The main difference between string
theories with target-space containing the (non-compact) line $\bb R^1$
or (compact) circle $S^1$ is the existence of winding modes in the
latter case, corresponding to the string wrapping around the circle
(see, for example, \cite{gsw,tdual}).
The major problem with introducing  winding modes in
{\sc tm} theory is the fact that the corresponding
worldsheet vertex operators must have different left and right
momenta. As we shall discuss in detail later, worldsheet vertex 
operators are represented in {\sc tm} thoery by Wilson lines 
of charged particles propagating between the left and right 
boundaries (which represent the left and right string worldsheets) 
of the topological membrane. 
The charge along one of these Wilson lines is interpreted as 
giving the momentum of the corresponding vertex operator. 
If the vertex operator has different left and right momenta,
then the charge along the Wilson line connecting left and right
membrane boundaries must change accordingly. 
So, in order to construct string winding modes, we need some process 
which leads to charge non-conservation in compact {\sc tmgt}.
Fortunately, precisely this type of process was discussed by 
Lee \cite{lee1992a}, who considered the effect of monopole-instantons
in compact  Chern-Simons gauge theory. The presence of 
monopole-instantons in compact $U(1)$ {\sc tmgt} was also discussed 
in \cite{compact} using a Hamiltonian approach, where it was also
found that monopole-instantons induce a phase transition in the bulk
matching precisely the {\sc bkt} phase transition \cite{bkt} on the 
string worldsheet \cite{bktstring}.
The purpose of this letter is to explicitly construct the string 
winding modes in {\sc tm} theory using the fact that
monopole-instantons lead to charge non-conservation in compact 
{\sc tmgt}. We interpret the amount of charge non-conservation 
as giving the momentum of the string winding mode and show that 
the resultant spectrum matches precisely the mass spectrum for 
string theory compactified on a circle. We then generalize our 
argument to include the case of compactification on a $D$-dimensional 
torus.
   
We first recall how string scattering amplitudes can be expressed
in terms of correlation functions of some 2d conformal field theory.
For simplicity, we consider closed bosonic strings in the critical
dimension which means that we may neglect 2d gravity. The
scattering amplitude has the general form:
\begin{equation}
A(p_1,\ldots , p_N)\sim \left\langle\prod_{i=1}^{N}
V_i(p_i^\mu)\right\rangle\, ,
\label{ampl}
\end{equation}
where $V_i(p_i^\mu)$ is the vertex operator corresponding to a 
particle of type $i$ with momentum $p_i^\mu$. Averaging is done
using the path integral:
\begin{equation}
\int{\cal D}X^\mu(\sigma)\exp\left\{-\frac{1}{4\pi\alpha^\prime}
\int\mbox{d}^2\!\sigma\,\eta^{\alpha\beta}\partial_\alpha X^\mu
\partial_\beta X_\mu\right\}\, ,  
\end{equation}
where $\alpha,\beta=1,2$ and $\mu=1,\ldots,26$. Compactification 
on the circle $S^1$ proceeds by identifying $X_{26}=X_{26}+2\pi\!R\,n$
where $R$ is the radius of $S^1$ and $n$ is the number of times the
string winds around the circle. The compact part of the string action 
becomes
\begin{equation}
S_{XY} = -\frac{R^2}{4\pi\alpha^\prime}\int\mbox{d}^2\!\sigma\,
(\partial_\alpha\theta)^2\, ,
\label{xy}
\end{equation}
where $\theta\in[0,2\pi)$. The resulting mass spectrum, obtained
from the on-shell condition 
$2(L_0+\tilde{L}_0-2)=0$ (see \cite{gsw} for details), is given by 
\begin{equation}
\alpha^\prime
M^2=-4+2(N_R+N_L)+m^2\frac{\alpha^\prime}{R^2}+n^2\frac{R^2}
{\alpha^\prime}
\ \ \ \ \mbox{with} \ \ \ \  N_L-N_R=nm.
\label{mass}
\end{equation}
$N_L$ and $N_R$ are the numbers of left- and right-moving excitations
and $m$ is an integer describing the allowed momentum eigenvalues in 
the compact direction. The last two terms in the spectrum 
give the contributions of the compact momentum and 
winding energy to the 25-dimensional mass. It easy to see that
there is a symmetry:
\begin{equation}
R\leftrightarrow \alpha^\prime/R \ \ \ \ \mbox{and} \ \ \ \ m\leftrightarrow
n\, ,
\label{tdual}
\end{equation}
which leaves the spectrum (\ref{mass}) invariant. This is the famous 
$T$-duality  of compactified string 
theory (for a review, see \cite{tdual}). 

\begin{figure}[htb]
\centerline{\psfig{figure=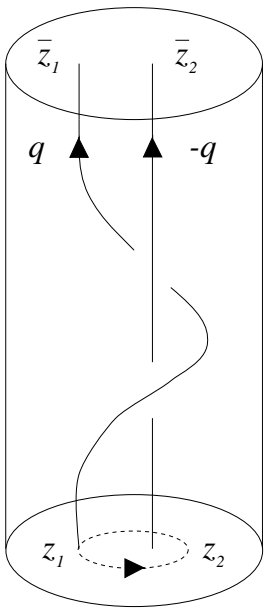,height=0.5\textwidth}}
\caption{String scattering amplitude as the correlator of 
Wilson lines in  the bulk {\sc tmgt}. Adiabatic rotation of 
the chiral vertex operators $V(z_1)$ and $V(z_2)$ corresponds 
to the linking of the two Wilson lines.}
\label{link}
\end{figure}

Before proceeding to show how the compactified string spectrum arises 
in {\sc tm} theory, we first recall how to obtain
string scattering amplitudes (\ref{ampl}) from topological membranes
\cite{tmtheory}. Using the normalizations
of \cite{heter1}, it is known that the abelian {\sc tmgt}
\begin{equation}
S_{\tiny TMGT}=-\frac{1}{4e^2}\int_{\cal M}F_{\mu\nu}
F^{\mu\nu} + \frac{k}{8\pi}\int_{\cal M}\epsilon^{\mu\nu\lambda}
A_\mu\partial_\nu A_\lambda\, ,
\label{tmgt}
\end{equation}
defined on a three-manifold $\cal M$, induces the chiral string action
\begin{equation}
S =\frac{k}{8\pi}\int_{\partial\cal M}\partial_z\theta\,
\partial_{\bar{z}}\theta
\label{bosonic}
\end{equation}
where $\theta$ is the pure gauge part of $A_\mu$ on the boundary 
$\partial{\cal M}$. When $\cal M$ is 
the filled cylinder depicted in Figure \ref{link}, its two 
boundaries represent the
left- and right-moving sectors of the string worldsheet.  
Worldsheet vertex operators can be constructed if we allow charged
matter (either bosonic or fermionic) to propagate inside the
topological membrane. The vertex operator
\begin{equation}
V_q(z,\bar{z}) = V_q(z)V_q(\bar{z})=e^{iq\theta(z)}e^{iq\theta(\bar{z})}
\label{vertex}
\end{equation} 
can be obtained from the bulk {\sc tmgt} as the open Wilson line
\begin{equation}
W_q[C]=\exp\left(iq\int_C A_\mu\,\mbox{$dx$}^\mu\right)
\label{wilson}
\end{equation} 
where the contour $C$ connects left and right boundaries.
Since the vector potential becomes pure gauge on the boundary, 
it is easy to see that the Wilson line (\ref{wilson}) coincides 
on the left and right boundaries with the holomorphic and 
anti-holomorphic parts of the vertex operator (\ref{vertex}), 
respectively. Moreover, the charge along the Wilson trajectory
is to be interpreted as giving the momentum of the corresponding
vertex operator. This suggests that the $N$-point function 
(\ref{ampl}) should be related to the correlator of $N$ Wilson lines 
in {\sc tmgt}. For the simplest case, $N=2$, the 
short-distance operator product expansion gives
\begin{equation}
\left\langle
V_q(z_1)V_{-q}(z_2)\right\rangle\sim(z_1-z_2)^{-2\Delta}\, ,
\label{ope}
\end{equation}
where $\Delta$ is the scaling dimension of the
chiral vertex operator $V_q(z)$. The corresponding
three-dimensional picture is that of a charged particle-antiparticle
pair propagating inside the membrane. The two-point function 
(\ref{ope}) should be related to the correlator of two Wilson 
lines in  {\sc tmgt} which, in the infrared (Chern-Simons) limit, 
is simply 
\begin{equation}
\left\langle W[C_1]W[C_2] \right\rangle = \exp \left\{-4\pi i\,
(q^2/k)\,\gamma[C_1,C_2]\right\}
\label{linking}
\end{equation} 
where $\gamma[C_1,C_2]$ is the linking number of the curves $C_1$
and $C_2$. The relationship between the two- and three-dimensional
correlation functions becomes clear if we adiabatically rotate 
$V(z_1)$ around $V(z_2)$. This induces a phase factor 
$e^{-4\pi i\Delta}$ in (\ref{ope}). 
A similar phase factor appears in (\ref{linking}) due to the 
linking of the two Wilson lines (see Figure \ref{link}). 
Since
\begin{equation}
\Delta=q^2/k
\label{scaling}
\end{equation}
is the transmuted spin of the charged particle due to 
its interaction with the Chern-Simons gauge field, this establishes
the equivalence between the anomalous spins in the two- and 
three-dimensional theories. Note that the spectrum of anomalous 
dimensions can also be obtained perturbatively from the 
{\sc tmgt} \cite{szaboET}.
Vertex operators representing higher-spin states have the
general form \cite{gsw}:
\begin{equation}
V_q(z,\bar{z})\sim\prod_j\left(\partial^j_z\theta\right)^{m_j}
\!e^{iq\theta(z)} 
\prod_k\left(\partial^k_{\bar{z}}\theta\right)^{m_k}
\!e^{iq\theta(\bar{z})}\, ,
\label{spinops}
\end{equation}
where the number of left- and right-moving excitations are 
given by $N_L=\sum_j jm_j$ and $N_R=\sum_k km_k$, respectively.
Since the vector potential $A_\mu$ is pure gauge on the
boundary, we can construct the pre-exponential (spin) factors 
in {\sc tm} theory using the boundary value of $A_\mu$ and 
its derivatives. For details, we refer the reader to \cite{tmtheory}.

We now turn our attention to the topological membrane description of 
string compactification on $S^1$. Comparing the boundary action
(\ref{bosonic}) with the compact part of the string action 
(\ref{xy}) shows that the compactification radius $R$ is related 
to the Chern-Simons coefficient $k$ by $k=4R^2/\alpha^\prime$.
Moreover, if $\theta\in [0,2\pi)$ then we must take
the gauge group in (\ref{tmgt}) to be compact $U(1)$. 
As we shall see, there is a crucial difference between
compact and non-compact {\sc tmgt} which enables us to construct 
string winding modes in the former theory. The difference
becomes apparent when we try to quantize the bulk theory. 
Quantizing in the $A_0=0$ gauge gives the equal-time commutation 
relations:
\begin{eqnarray}
[E_i(x),E_j(y)] &=& -\mbox{$\frac{ik}{4\pi}$}\epsilon_{ij}\delta(x-y) 
\nonumber\\  
\!\,[E_i(x),B(y)]   &=& -i\epsilon_{ij}\partial_j\delta(x-y) 
\end{eqnarray}
where $E_i=\Pi_i-k/8\pi\epsilon_{ij}A^j$ and $B=\epsilon^{ij}
\partial_iA_j$. The classical Gauss law is
\begin{equation}
\partial_i E^i +\frac{k}{4\pi}B = \rho\, ,
\label{gauss}
\end{equation}
where $\rho$ is the charge density of the charged matter.
As usual, Gauss' law generates time independent gauge transformations.
The elements of the gauge group are the operators:
\begin{equation}
U=\exp\left\{-i\int\mbox{d}^2z\,\lambda(z)\Big(\partial_iE^i+
\mbox{$\frac{k}{4\pi}$}B - \rho\Big)\right\}\, ,
\label{group}
\end{equation}
with 
$UA_i\,U^{-1}=A_i+\partial_i
\lambda$. The physical Hilbert space of the theory contains only
those states which are invariant under the action of $U$, i.e. 
$U|\Psi\rangle = |\Psi\rangle$. If the gauge group is compact,
however, we must be more careful. As described in \cite{compact}
(see also \cite{luscher1989a}), we must also include in the gauge group
operators of the form (\ref{group}) where $\lambda(z)$ is now the
angle in the complex plane. When $z$ is restricted to a simply
connected region not containing the origin, $\lambda(z)$ is 
differentiable and the Cauchy-Riemann equations give
\begin{equation}
\partial_i\lambda(z)=-\epsilon_{ij}\,\partial_j\ln |z|\, .
\end{equation}
So in the compact theory we must further restrict the Hilbert space 
to eigenstates of the operator
\begin{equation}
V(y)=\exp\left\{-i\int\mbox{d}^2z\,\Big(E^i+\mbox{$\frac{k}{4\pi}$}
\epsilon^{ij}A_j
\Big)\,\epsilon_{ik}\,\partial_k\ln |y-z| - \lambda(y-z)\rho\right\}\, .
\label{vortex}
\end{equation}
Using the identity $\partial_k\partial_k\ln |z| = 2\pi \delta(z)$,
it is straightforward to show that 
\begin{equation}
[B(x),V^n(y)]=2\pi n\,\delta(x-y)\,V^n(y)\, ,
\label{flux}
\end{equation}
and so $V^n(y)$ creates a pointlike magnetic vortex with flux 
$2\pi n$. Integrating the Gauss law (\ref{gauss}) and noting 
that the electric field falls off exponentially (the photon in 
(\ref{tmgt}) is massive), shows that the operator $V^n(y)$ 
also creates electric charge
\begin{equation}
\Delta Q=nk/2\, .
\label{viol}
\end{equation}
This change of charge/flux is, however, unobservable far from the 
vortex because local observables such as the electric field fall 
off exponentially and the Aharonov-Bohm phase is unity. Therefore 
$V(y)$ is the operator for a monopole-instanton \cite{lee1992a} 
which interpolates between topologically inequivalent vacua. 

Given that the compact theory contains monopole-instantons which
carry quantized charge/flux, we ask what is spectrum of allowed
charge in the compact {\sc tmgt}? Since the charge along a Wilson
trajectory specifies the momentum of the corresponding worldsheet
vertex operator, finding the spectrum of charge is equivalent to
finding the spectrum of allowed string momenta. To find the spectrum,
we first write the gauge group generators in (\ref{group}) in terms of
the canonical momenta $\Pi^i=E^i+k/8\pi\,\epsilon^{ij}A_j$. Using
the functional Schr\"odinger representation ($\Pi^i=i\,\delta/\delta 
A_i$) it is easy to see that physical states acquire a non-trivial
phase under gauge transformations:
\begin{equation}
U\Psi[A_i]=\exp\left\{-i\int\lambda\Big(\mbox{$\frac{k}{8\pi}$}B
- \rho\Big)\right\}\,\Psi[A_i+\partial_i\lambda]\, . 
\label{phase}
\end{equation}
In the compact theory, $\lambda$ is defined modulo $2\pi$ and
the phase factor in (\ref{phase}) must be invariant under  
$\lambda\rightarrow \lambda+2\pi$. This implies that the charge
$Q=\int\mbox{d}^2\!z\,\rho$ must be quantized according to
\begin{equation}
Q = m + \mbox{$\frac{k}{8\pi}$}\!\int\!B\, ,
\label{quant}
\end{equation}
where $m$ is an integer. If $k=0$ then we recover the more
familiar charge quantization condition of compact electrodynamics.
The extra term in (\ref{quant}) is due to the monopole-instanton
background which, according to  (\ref{flux}), carries magnetic flux 
$2\pi n$. Hence the spectrum of allowed charge (momenta) in the
compact theory  is
\begin{equation}
Q=m +nk/4\, .
\end{equation}

\begin{figure}[htb]
\centerline{\psfig{figure=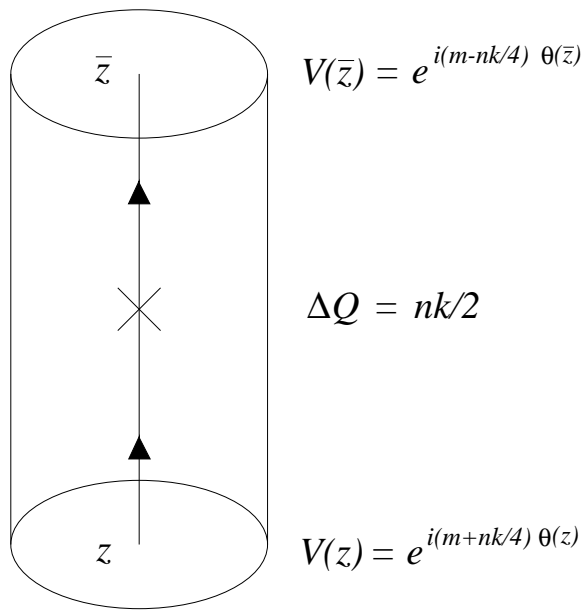,height=0.5\textwidth}}
\caption{Construction of a string winding mode from compact
{\sc tmgt}. A monopole-instanton transition causes charge
non-conservation along the Wilson trajectory connecting left 
and right boundaries (worldsheets). The difference between  
left and right boundary charges (momenta) is interpreted
as a string winding mode.}
\label{instanton}
\end{figure}

To see how the presence of monopole-instantons in the compact 
{\sc tmgt} generates the mass spectrum (\ref{mass}), consider the 
process illustrated in Figure \ref{instanton}. 
A particle of charge $Q=m+nk/4$ is inserted
on the left boundary which then propagates through the bulk. The charged
particle interacts with the monopole-instanton background which,
according to (\ref{viol}), itself carries charge $nk/2$. The effect 
of the monopole-instanton is to change the charge of the particle, 
creating a new state of charge $Q=m-nk/4$, which propagates
to the right boundary. Using (\ref{scaling}), the scaling dimension
of the corresponding worldsheet vertex operator is 
\begin{equation}
\Delta+\bar{\Delta}  = \frac{(m+nk/4)^2}{k} + \frac{(m-nk/4)^2}{k}\, .
\end{equation}
Substituting $k=4R^2/\alpha^\prime$ we obtain
\begin{equation}
2(\Delta +\bar{\Delta})= m^2\frac{\alpha^\prime}{R^2}
+n^2\frac{R^2}{\alpha^\prime}\, ,
\label{tmgtspectrum}
\end{equation}
which matches precisely the last two terms 
of (\ref{mass}), i.e. the contribution of the compact momentum and
winding energy to the 25-dimensional mass spectrum. The difference 
between the number of left- and right-moving excitations is given
by
\begin{equation}
N_L-N_R=\Delta-\bar{\Delta}=nm\, ,
\label{spin}
\end{equation}
which also agrees with (\ref{mass}). Since $\Delta$ is the induced 
spin of the charged particle due to its interaction with the 
Chern-Simons gauge field, (\ref{spin}) implies that the 
angular momentum  of the particle is not conserved as it
propagates from left to right boundary. However, the total
angular momentum in the membrane {\em is} conserved due to photon
emission in the bulk, which can be seen as follows. Since
$N_L\neq N_R$, then the vertex operator (\ref{spinops})
corresponding to the string winding mode must have a different
number of left and right pre-exponential (spin) factors. As
discussed earlier, these spin factors are constructed in {\sc tm}
theory using the boundary value of $A_\mu$ and its derivatives.
Hence, the three-dimensional interpretation of (\ref{spin}) is 
that there must be a different number of photon operators on
the left and right boundaries or, equivalently, that the charged
particle must emit photons as it propagates between boundaries,
thereby conserving the total angular momentum in the bulk.
We hope to give the precise details of this process in a future 
paper, including the possibility that the emitted photons may 
themselves interact with the monopole-instanton background.
 
We now show how the above analysis generalizes to spacetime with 
higher compact dimension. Suppose the compactified
space is a $D$ dimensional torus by identifying
$X^I = X^I + 2\pi L^I$, where $I=1\dots D$. 
The compact part of the string action
now takes the form:
\begin{equation}
S = -\frac{1}{4\pi\alpha^\prime}\int\mbox{d}^2\!\sigma\,
\left(G_{IJ}\,\partial_\alpha\theta^I\partial_\alpha\theta^J
+\epsilon^{\alpha\beta}B_{IJ}\,\partial_\alpha\theta^I\partial_\beta
\theta^J\right)\, , 
\label{torusaction}
\end{equation}
where $G_{IJ}$ and $B_{IJ}$ are the graviton and antisymmetric 
tensor condensates. The allowed left- and right-moving
momenta are (see, for example, Eq.\,(10) of \cite{narainET1987})
\begin{eqnarray}
P_I &=& m_I-G_{IJ}\,n^J/\alpha^\prime-B_{IJ}\,n^J/\alpha^\prime
\nonumber\\
\bar{P}_I &=& m_I+G_{IJ}\,n^J/\alpha^\prime-B_{IJ}\,n^J/\alpha^\prime\, .
\label{momenta}
\end{eqnarray}
The momentum associated with the winding mode is just the difference
between the left and and right momenta, namely
\begin{equation}
\Delta P_I = 2G_{IJ}\,n^J/\alpha^\prime\, .
\label{windingmom}
\end{equation}
The resultant mass spectrum is
\begin{equation}
\alpha^\prime M^2=-4+2(N_R+N_L)+\alpha^\prime\,m_I\,G^{IJ}m_J
+\frac{1}{\alpha^\prime}\,n^I (G-BG^{-1}B)_{IJ}\,n^J + 2n^IB_{IK}G^{KJ}
m_{J}\, ,
\label{torusmass}
\end{equation}
which is invariant under
\begin{equation}
\frac{1}{\alpha^\prime}(G\pm B)_{IJ}\rightarrow \alpha^\prime
(G\pm B)_{IJ}^{-1} = \alpha^\prime(G\pm B)^{IJ}
\label{torusdual}
\end{equation}
with the simultaneous exchange of $m_I\leftrightarrow n^J$.

To obtain the topological membrane description of the above
toroidal compactification, we consider the abelian 
{\sc tmgt} with $D$ copies of $U(1)$:
\begin{equation}
S=-\frac{1}{4e^2}\int_{\cal M}F_{\mu\nu}^I
F^{\mu\nu}_I + \frac{K_{IJ}}{8\pi}\int_{\cal M}\epsilon^{\mu\nu\lambda}
A^I_\mu\partial_\nu A^J_\lambda\, , 
\label{torustmgt}
\end{equation}
where $K_{IJ}$ is some non-degenerate matrix (not necessarily 
symmetric). The induced chiral action
on the boundary is  
\begin{equation}
S =\frac{K_{IJ}}{8\pi}\int_{\partial\cal M}\partial_+\theta^I\,
\partial_{-}\theta^J\, ,
\end{equation}
which matches the string action (\ref{torusaction}) if $K_{IJ}=4(G_{IJ}
+B_{IJ})/\alpha^\prime$. The scaling dimension (\ref{scaling}) 
generalizes to 
\begin{equation}
\Delta = K^{IJ}Q_I Q_J\, ,
\end{equation}
where $K^{IJ}=K_{IJ}^{-1}$.
Quantization in the $A_0^I = 0$ gauge
proceeds analogously to the $U(1)$ case except for the following
modifications. The equal-time commutation relations are:
\begin{equation}
[\Pi^i_I(x),A^j_J(y)] = -i\,\delta^{ij}\delta_{IJ}\delta(x-y) 
\end{equation}
where $\Pi_I^i=E_I^i+(K_{JI}/8\pi)\,\epsilon^{ij}A_j^J$.
The Gauss law (\ref{gauss}) generalizes to
\begin{equation}
\partial_i E^i_I +\frac{\tilde{G}_{IJ}}{4\pi}\,B^J = \rho_I\, ,
\label{torusgauss}
\end{equation}
where $\tilde{G}_{IJ}=4G_{IJ}/\alpha^\prime$ is the symmetric part of 
$K_{IJ}$. The
elements of the gauge group are
\begin{equation}
U = \exp\left\{-i\int\lambda^I\Big(\partial_i E^i_I+
\tilde{G}_{IJ}/4\pi\,B^J - \rho_I\Big)\right\}\, ,
\label{torusgroup}
\end{equation}
where $\lambda^I$ parameterizes each compact $U(1)$ factor.
Associated with each of these $U(1)$ factors is the 
monopole-instanton operator
\begin{equation}
V_I(y)=\exp\left\{-i\int\mbox{d}^2z\,\Big(E^i_I+\tilde{G}_{IJ}/4\pi\,
\epsilon^{ij}A_j^J\Big)\,\epsilon_{ik}\,\partial_k\ln |y-z| 
- \lambda_I(y-z)\rho\right\}\, ,\nonumber
\label{torusvortex}
\end{equation}
with
\begin{equation}
[B_I(x),V_J^n(y)]=2\pi n\,\delta_{IJ}\,\delta(x-y)\,V^n(y)\, .
\label{torusflux}
\end{equation}
Hence the flux carried by each monopole-instanton is $2\pi\,n^J$.
Integrating the Gauss law (\ref{torusgauss}) shows that the 
monopole-instanton also carries charge 
\begin{equation}
\Delta Q_I=\tilde{G}_{IJ}\,n^J/2 = 2G_{IJ}\,n^J/\alpha^\prime\, .
\label{toruscharge}
\end{equation}
Just like the $U(1)$ case, the charge non-conservation induced
by the monopole-instanton matches the string winding 
momenta (\ref{windingmom}). However, now we encounter an apparent
paradox: how does the bulk {\sc tmgt} account for the extra
terms in (\ref{momenta}) involving the antisymmetric tensor
$B_{IJ}$? The resolution of this paradox emerges if we write
the gauge group generators in (\ref{torusgroup}) in terms of
the canonical momenta 
$\Pi_I^i=E_I^i+(K_{JI}/8\pi)\,\epsilon^{ij}A_j^J$. Using the
functional Schr\"{o}dinger representation ($\Pi_I^i=i\,\delta/
\delta A^I_i$), it is clear that the phase acquired by physical 
states under a gauge transformation  
\begin{equation}
U\Psi[A_i^I]=\exp\left\{-i\int\lambda^I\Big(K_{IJ}/8\pi\,B^J
- \rho_I\Big)\right\}\,\Psi[A_i^I+\partial_i\lambda^I]\, . 
\label{states}
\end{equation}
involves the full matrix $K_{IJ}$. Requiring that the phase factor 
in (\ref{states}) be invariant under $\lambda^I\rightarrow \lambda^I
+2\pi$ gives the quantization condition
\begin{equation}
Q_I = m_I - \frac{K_{IJ}}{4}n^J\, ,
\end{equation}
where we have used the fact that the monopole-instanton associated 
with each compact $U(1)$ factor carries magnetic flux $2\pi\,n^J$.
We are now in a position to show how the spectrum of left and
right momenta (\ref{momenta}) arises from the following process 
in the bulk. On the left boundary we insert the
charge
\begin{equation}
Q_I = m_I - \frac{\tilde{G}_{IJ}}{4}n^J -
\frac{\tilde{B}_{IJ}}{4}n^J\, ,
\end{equation}
where $\tilde{B}_{IJ}=4B_{IJ}/\alpha^\prime$ is the antisymmetric 
part of $K_{IJ}$. The
charged matter then propagates through the bulk where, according
to (\ref{toruscharge}), interaction
with the monopole-instanton background changes the charge by 
$\tilde{G}_{IJ}\,n^J/2$. This creates a new state with charge
\begin{equation}
\bar{Q}_I = m_I + \frac{\tilde{G}_{IJ}}{4}n^J - \frac{\tilde{B}_{IJ}}{4}n^J
\end{equation} 
which propagates to the right boundary. Substituting
$K_{IJ}=4(G_{IJ}+B_{IJ})/\alpha^\prime$ gives
\begin{eqnarray}
Q_I &=& m_I-G_{IJ}\,n^J/\alpha^\prime-B_{IJ}\,n^J/\alpha^\prime
\nonumber\\
\bar{Q}_I &=& m_I+G_{IJ}\,n^J/\alpha^\prime-B_{IJ}\,n^J/\alpha^\prime\, ,
\end{eqnarray}
which matches precisely the spectrum of left and right momenta 
(\ref{momenta}). As a final check, the scaling dimension of the 
corresponding worldsheet vertex operator is
\begin{eqnarray}
2(\Delta+\bar{\Delta}) &=& 2K^{IJ}(Q_IQ_J+\bar{Q}_I\bar{Q}_J)
\nonumber\\
&=& \alpha^\prime\,m_I\,G^{IJ}m_J
+\frac{1}{\alpha^\prime}\,n^I (G-BG^{-1}B)_{IJ}\,n^J + 2n^IB_{IK}G^{KJ}
m_{J}\, ,
\label{torusspec}
\end{eqnarray}
which indeed matches the last three terms of (\ref{torusmass}),
i.e. the contribution of the compact momenta and winding energy to the
$(26-D)$-dimensional mass. 

It is straightforward to show that the compactified spectrum 
(\ref{torusspec}) exhibits the \mbox{$T$-duality} (\ref{torusdual}). 
The question then arises: what
is the meaning of this $T$-duality from the three-dimensional
point of view? This question was addressed in \cite{tmdual}
where it was shown that the three-dimensional analogue of 
$R\rightarrow \alpha^\prime/R$ duality (and more generally 
$K_{IJ}\rightarrow K_{IJ}^{-1}$ duality) is the duality between 
large and small scales in the corresponding {\sc tmgt} with the 
spontaneous breaking of gauge symmetry. This duality was shown to
be a consequence of the equivalence between {\sc tmgt} and 
Chern-Simons gauge theory with a Proca mass term.
In \cite{tmdual} however, it was not at all clear how to describe 
the string winding modes in three-dimensional terms. Now that 
we have shown how string winding modes correspond to
monopole-instantons in the bulk {\sc tmgt},  
and motivated by the fact that $T$-duality relates winding modes 
in one string theory to momentum modes in the dual string theory,
we suggest that the corresponding bulk duality should relate
monopole-instantons (i.e. a non-perturbative bulk effect) 
in the original {\sc tmgt} to a charge conserving (perturbative) 
process in the dual gauge theory. We hope to study the precise nature 
of this duality in future work. 

We should point out that the three-dimensional analogue of 
$R\rightarrow \alpha^\prime/R$ duality
has also been discussed \cite{bal} in the context of the quantum
Hall effect which, itself, can be described in terms of a Chern-Simons
gauge theory. Moreover, it is interesting to note that the picture of
charge transport between membrane boundaries presented in this 
letter also has a close counterpart in the quantum Hall effect.
Indeed, tunneling between edge states has proved useful in 
classifying the internal topological orders in the quantum
Hall effect \cite{wen} and has been supported by many experiments.
Given that it remains an open question whether the quantum Hall
effect is described by compact or non-compact Chern-Simons theory,
and motivated by the connection with  string winding modes
described in this letter, it would be extremely interesting to search 
for some signal of charge non-conservation in a quantum Hall tunneling
experiment.

In closing, we would like to mention a surprising (and possibly
important) prediction that {\sc tm} theory makes in relation
to heterotic string theory. As described in \cite{heter1}, 
the different left and right sectors of the heterotic string can 
be obtained in {\sc tm} theory by imposing certain boundary 
conditions on the bulk gauge fields. In particular, on
one of the membrane boundaries we must fix both 
$A_z=A_{\bar z}=0$ ({\sc n} boundary conditions in the
notation of \cite{heter1}) which, in turn forces the
magnetic field $B=0$ on that boundary. The Gauss law
(\ref{gauss}) then implies that there can be no 
charged particles on that particular boundary ---  and this is 
a problem if we want to construct vertex operators from 
Wilson lines of charged particles propagating between 
left and right membrane boundaries. Clearly the problem 
is solved if the theory is compact, whereby a monopole-instanton 
transition can change the charge to zero along a Wilson trajectory. 
However, in the non-compact {\sc tmgt} there is no known physical 
mechanism for changing the charge in the bulk and so this suggests 
that each $U(1)$ factor must be compact. Since each $U(1)$ factor 
corresponds to a particular space-time co-ordinate of the induced
string theory, we must therefore conclude that if heterotic strings 
are described by topological membranes then each and every space-time 
dimension must be compact! We emphasize that the radii of the four
Minkowski space-time dimensions may still be large (this is a question
of dynamics) but, nevertheless, a consistent {\sc tm} theory of 
heterotic strings requires that they must be compact.

To conclude, we have shown how to construct string winding modes from
a process of charge non-conservation in {\sc tm} theory. Winding modes
are the simplest example of solitonic states in string theory and we 
have shown that they arise in {\sc tm} theory as a result of
a non-perturbative effect (monopole-instantons) in the membrane 
bulk. We speculate that other (if not all) soliton states in string
theory can be described in {\sc tm} theory as some non-perturbative
bulk process.   

We thank Richard Szabo and Michael Birkel for
helpful discussions.
One of us ({\sc l.c.}) gratefully acknowledges financial support
from the University of Canterbury, New Zealand.

\end{document}